Fermionic Critical Fluctuations: Potential Driver of Strange Metallicity and Violation of the Wiedemann-Franz Law in YbRh$_2$Si$_2$

Frank Steglich[1,2]

[1]Max Planck Institute for Chemical Physics of Solids (MPI CPfS), 01187 Dresden, Germany
[2]Center for Correlated Matter (CCM) and School of Physics, Zhejiang University, Hangzhou 310058, China

Abstract

Results of combined thermal and electrical transport measurements through the magnetic field-induced quantum critical point in the heavy-fermion compound YbRh$_2$Si$_2$ are revisited to explore the relationship between the strange-metal behavior, observed in both the electrical and electronic thermal resistivity, and the violation of the Wiedemann-Franz law in the zero-temperature limit. A new type of inelastic scattering center for the charge and heat carriers has been detected and ascribed to the small-to-large Fermi-surface fluctuations. These are operating in the vicinity of and at the Kondo-destroying quantum critical point as fermionic quantum critical fluctuations and are considered the primary driver of the strange-metal behavior and the violation of the Wiedemann-Franz law.

I. Mott-Type Quantum Critical Point

Quantum criticality in strongly correlated electron systems continues to be an exciting research field in condensed-matter physics. Much insight into this topic has been achieved on antiferromagnetic (AF) heavy-fermion compounds [1,2]. The conventional itinerant spin-density-wave-type instability at zero temperature, explored by the theorists already many years ago [3-5] was studied in detail for the heavy-fermion metal $CeCu_2Si_2$ [6]. While in this case the charge carriers, i.e., propagating Kondo singlets, stay intact at the AF instability, they apparently break apart at the AF quantum critical point (QCP) investigated in the heavy-fermion metals $CeCu_{5.9}Au_{0.1}$ [7,8] and $CeRhIn_5$ [9-11]. As the so-called Kondo-breakdown instability in the two latter compounds involves a 4f-electron localization/delocalization mechanism, it is frequently labeled a "partial Mott" or "Mott-type" quantum phase transition. This kind of unconventional local QCP [12,13] gives rise to "strange metallicity", which is of high interest in contemporary research on strongly correlated materials [14-18] and has been discussed in connection to astrophysics [19]. Strange-metal behavior is illustrated by a logarithmic divergence of the Sommerfeld coefficient of the electronic specific heat as well as a linear temperature dependence of the electrical resistivity, $\Delta\rho(T) = [\rho(T) - \rho_0] = aT$ ($\rho_0$: residual resistivity) [2,20,21]. A Kondo-destroying QCP has also been established for the canonical heavy-fermion metal $YbRh_2Si_2$ under magnetic-field tuning at a very low critical field [$B_N = 0.06$ T ($\perp$ c-axis)], where AF order smoothly disappears [22,2], as shown in Fig. 1 [23]. At $B = B_N$, the crossover temperature $T_{FL}(B)$ also vanishes, below which heavy Fermi-liquid (FL) behavior emerges on the paramagnetic side of the phase diagram.

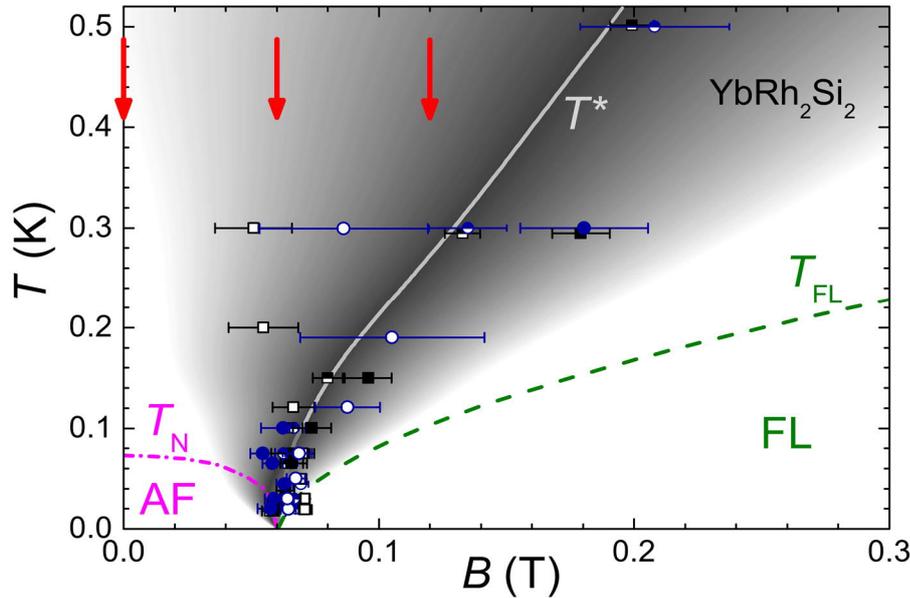

Fig. 1. Temperature (T) – magnetic field (B) phase diagram for YbRh$_2$Si$_2$ displaying a Kondo-breakdown quantum critical point at T = 0, $B_N$ = 0.06 T ($\perp$ c-axis), where the following coincide: (i) the phase-transition line $T_N(B)$ of the antiferromagnetically ordered phase at low fields, (ii) the crossover line $T_{FL}(B)$ to the low-T heavy Fermi-liquid phase at elevated fields and (iii) $T^*(B)$, indicating a Mott-type crossover from small-to-large Fermi surface. $T^*(B)$ was determined from the midpoints of thermally broadened jumps in the isothermal field dependences of magneto-transport results. The corresponding crossover widths are indicated by the horizontal bars. The grey scale visualizes the slope of isothermal magneto-resistance as a function of field. Red arrows indicate B = 0, $B_N$ and $2B_N$, respectively. Reproduced from Schuberth et al. [23].

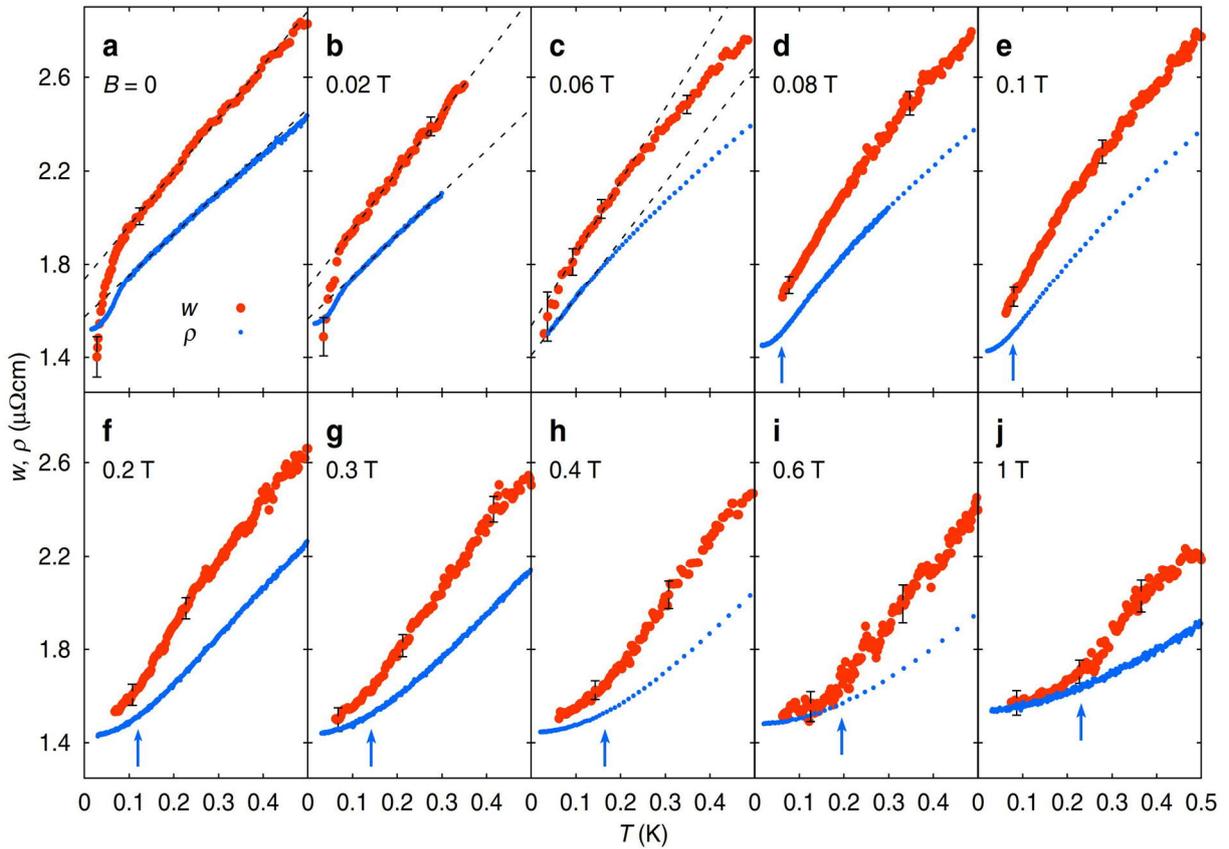

Fig. 2. Combined electrical and thermal transport though the quantum critical field $B_N = 0.06$ T ($\perp$ c-axis) in single-crystalline $YbRh_2Si_2$. Electrical and thermal resistivities, $\rho(T)$ and $w(T) = L_0T/\kappa(T)$ [with same units], are displayed as a function of temperature for $T \leq 0.5$ K at $B = 0$, $B = B_N$ and varying magnetic fields below and above $B_N$. $\kappa(T)$ is the measured thermal conductivity, and $L_0 = (1/3)(\pi k_B/e)^2$ is Sommerfeld's constant. At $B = B_N$, strange-metal behavior is observed only for $T < 0.3$ K in $w(T)$ and $< 0.2$ K in $\rho(T)$, indicating a high quality of this single crystal, see [2]. In contrast, for a slightly Ge-substituted single crystal with ten times larger residual resistivity, strange-metal behavior was registered over three decades in T up to 10 K [22]. Reproduced from Steglich et al. [25].

In addition, another line, T*(B), starts at B = $B_N$ with an almost infinite slope, indicating the crossover from a small Fermi surface at low fields to a larger one at elevated fields [22,2]. This crossover continuously broadens with increasing field, as indicated in Fig. 1.

II. Strange-Metal Behavior in Thermal and Electrical Transport: Violation of the Wiedemann-Franz (WF) Law

The low-temperature electrical and thermal resistivities, $\rho(T)$ and $w(T) = L_0 T/\kappa(T)$, of YbRh$_2$Si$_2$ were determined in the field range 0 - 1 T [24]. Here, $\kappa(T)$ is the thermal conductivity as measured, and $L_0 = (\pi k_B)^2/3e^2$ is Sommerfeld's constant. As shown in Fig. 2 [25], at elevated temperatures (T > 0.1 K) the thermal resistivity is larger than its electrical counterpart, with both displaying linear-in-T behavior in a certain T-window. At very low fields, an extra magnon-derived contribution $\kappa_m(T)$ appears in the thermal conductivity below T ≈ 0.1 K. $\kappa_m(T)$ is also observed at the critical field $B_N$ = 0.06 T, where it originates in short-lived magnon ("paramagnon") excitations. A small contribution by $\kappa_m(T)$ appears to exist even up to B = 0.1 T. For B ≥ 0.2 T, $\Delta w(T) = [w(T) - w_0] = A'T^2$ at sufficiently low temperatures, while for both B = 0 and 0.02 T the electrical resistivity $\rho(T)$ shows a kink at and a FL-type quadratic temperature dependence [$\Delta\rho(T) = AT^2$] below the Néel temperature $T_N$. At $B_N$ = 0.06 T, $\rho$ is linear in T down to 0.01 K, the lowest accessible temperature [22]. Recently, $\Delta\rho(T) = [\rho(T) - \rho_0] = aT$ has been observed at temperatures as low as 0.001 K [26]. This suggests that, exactly at the critical field $B_N$, YbRh$_2$Si$_2$ is a strange metal down to T = 0. At B ≥ 0.08 T, the low-T electrical resistivity displays a strong $T^2$-dependence [24,25], which proves the existence of a heavy Fermi liquid.

Using the Lorenz number $L(T) = \rho(T)\kappa(T)/T$, the Lorenz ratio is defined by $L(T)/L_0 = \rho(T)/w(T)$. This ratio was determined for YbRh$_2$Si$_2$ by Pourret et al. [27] in an extended T-window for B = 0 and various applied fields at and above the critical field that here amounts to $B_N \approx 0.07$ T. As seen in Fig. 3, the results below $T \approx 0.1$ K for both $B = 0$ and $B = B_N$ display a pronounced upturn, a tiny remnant of which being recognized up to $B = 0.17$ T. This upturn is proportional to the bosonic conductivity $\kappa_m(T)$, which vanishes at $T = 0$. Therefore, the extra contribution to $L(T)/L_0$ must reach a maximum before disappearing at $T = 0$. In the heavy FL phase at $B = 0$ and all field values $B < B_N$, the low-$T$ electrical and electronic thermal resistivities $\rho(T)$ and $w_{el}(T)$, respectively, follow a $T^2$-dependence. Here, the WF law, which holds only for purely elastic scattering processes, must be valid in the zero-temperature limit. This implies that $L(0) = L_0$ and $\rho_0 = w_0$. The same is true for all field values $B > B_N$. However, at $B = B_N$, the Lorenz ratio amounts to $L(0)/L_0 \approx 0.9$ at $T = 0$, where $\kappa_m(T) = 0$. At this critical field, $\Delta\rho(T) = aT$ and, most likely, also $\Delta w_{el}(T) = a'T$ display strange-metal behavior to the lowest accessible temperatures. Both are of disparate size in that (i) the slope $a'$ is larger than its counterpart $a$, and (ii) the residual electronic thermal resistivity $w_{el,0}$ exceeds the residual electrical resistivity $\rho_0$ by about 10 %. This clearly indicates that the WF law is violated exactly at the Kondo- destroying QCP. A similar observation was reported for YbAgGe [28]. For a detailed discussion of the contrasting claims by other groups concerning $YbRh_2Si_2$, see Ref. [23].

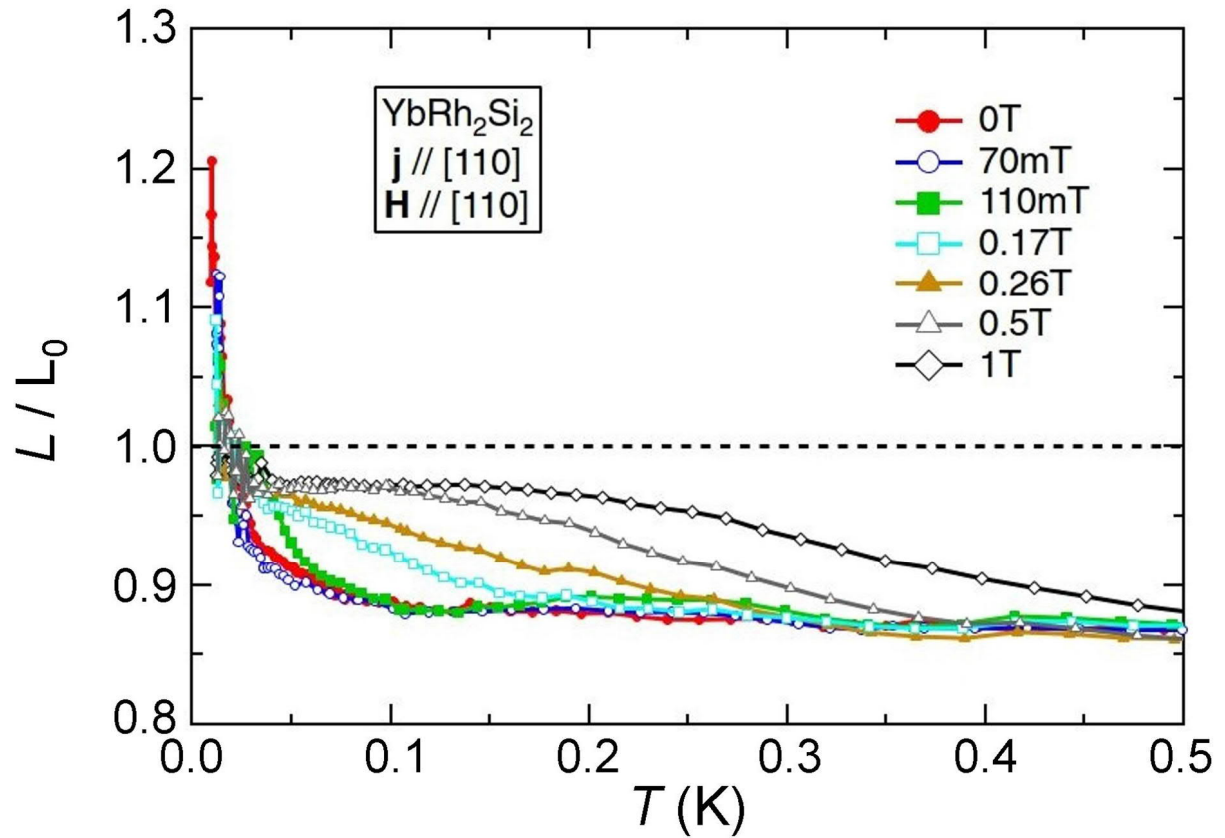

Fig. 3. Lorenz ratio $L(T)/L_0 = \rho(T)/w(T)$ for an YbRh$_2$Si$_2$ single crystal (with QCP at T = 0, $B_N$ ≈ 0.07 T) for B = 0, B = $B_N$ and varying fields B > $B_N$ in an extended temperature window, 0.08 K ≤ T ≤ 0.5 K. Reproduced with permission from Pourret et al. [27], The Physical Society of Japan Copyright (2014).

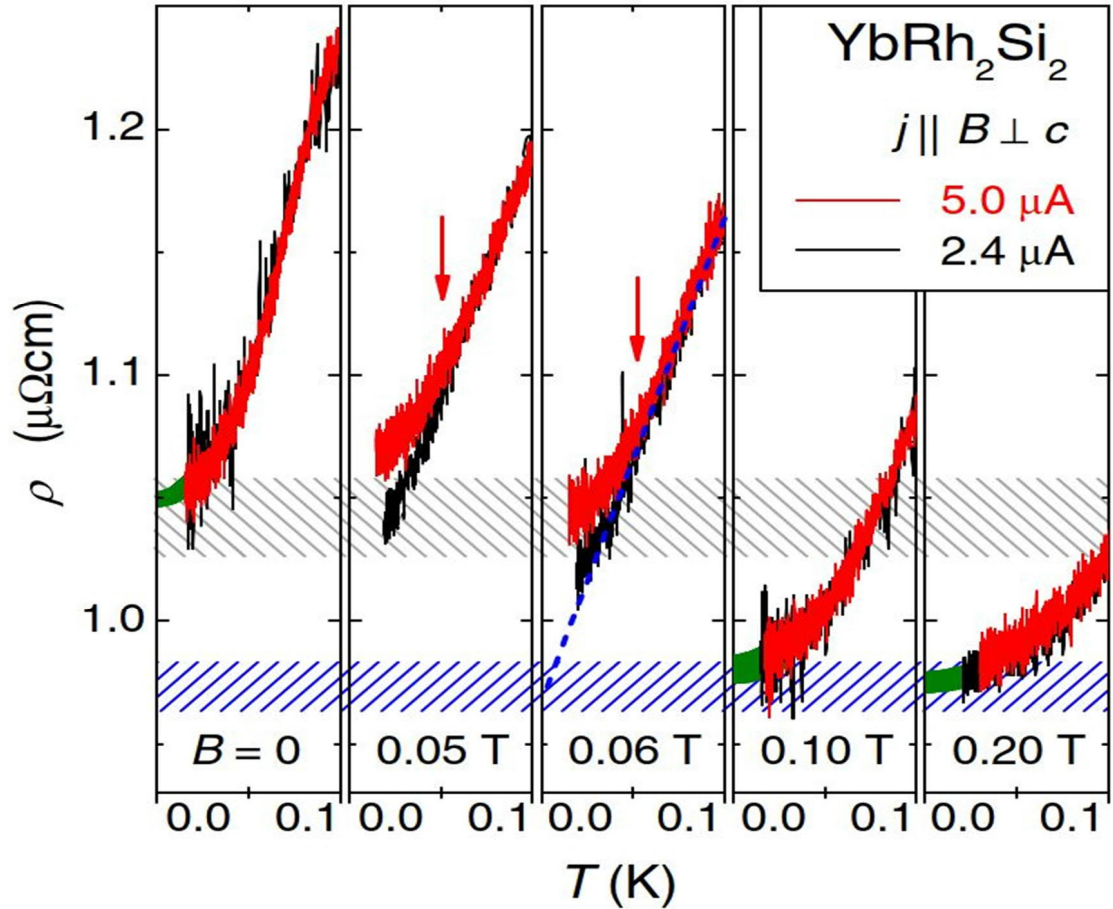

Fig. 4. $\rho$ vs $T$ at varying magnetic fields ($B$) for an $YbRh_2Si_2$ single crystal ($B_N = 0.059$ T). Black (small current $j$): no heating. At $B = 0$, 0.1 T and 0.2T, $\rho(T)$ shows a $T^2$-dependence at low temperatures. For $B = 0.05$ T ($\leq B_N$) and 0.06 T ($\geq B_N$), $\rho(T)$ is linear in $T$ to the lowest accessible temperature. The residual resistivity $\rho_0$ at $B = 0$ as well as 0.05 T (grey shading) is substantially higher than that at $B > B_N$ (blue shading). Extrapolations to $T = 0$ at/far away from $B_N$ are shown in blue/green. The abrupt change in $\rho_0$ indicates a reduced carrier density at low fields. Red (large $j$): for both $B = 0.05$ T and 0.06 T, $\Delta\rho(T)$ deviates from the linear-in-$T$ dependence (red arrows) and becomes proportional to $T^2$ at low temperatures. The observed heating on the approach of the QCP under large current conditions is owing to additional inelastic scatterings discussed in the text. Reproduced from Steglich et al. [25].

The apparent violation of the WF law in YbRh$_2$Si$_2$ was convincingly demonstrated by measuring the temperature dependence of the low-T electrical resistivity for two different currents (j = 2.4 and 5.0 µA, respectively) both near and far from the critical field $B_N$ = 0.059 T [25], specifically at B = 0.05 and 0.06 T as well as B = 0, 0.1 and 0.2 T (Fig. 4). In the vicinity of $B_N$, at B = 0.05 and 0.06 T, the large current of 5.0 µA causes the sample to be heated up at sufficiently low temperatures (red arrows), resulting in an enhanced $\rho$(T) that follows a FL-type quadratic T-dependence. Under the low current of 2.4 µA, the resistivity is smaller, exhibiting a linear T-dependence down to the lowest accessible temperature. In addition, no heating is observed apart from $B_N$, neither at B = 0 nor at B = 0.1 and 0.2 T. A possible explanation for this heating effect involves the strong reduction of the low-T electronic thermal conductivity $\kappa_{el}(T)$ in the vicinity of B = $B_N$ which overcompensates, under the large current (like under an increasing field, cf. Figs. 2d and e), a substantially weakened (para)magnon contribution $\kappa_m(T)$. As a result, the Joule heat produced by the large current will be prevented from being efficiently dissipated. Fig. 4 also shows a jump of the residual resistivity $\rho_0$ at or close to B = $B_N$, illustrating an abrupt increase in charge-carrier density with increasing field, as is indeed expected at the Kondo-destroying QCP [12,13,25].

To further investigate the apparent strange metallic ground-state of YbRh$_2$Si$_2$ and the simultaneous violation of the WF law at its QCP, the isothermal field tuning of $L(B)/L_0$ has been explored at temperatures T ≥ 0.1 K - within a temperature range where, owing to the absence of any significant (para)magnon excitations, the thermal transport is predominantly electronic [24]. The results presented in Fig. 5 were obtained from the temperature traces shown in Fig. 2. The Lorenz ratio is found to be smaller than one across the entire (T, B) range displayed in Fig. 5. This indicates dominating inelastic scattering processes, notably electron-electron and electron-spin fluctuation scatterings, both of which disappear in the zero-temperature limit because of phase-space arguments in the former and the bosonic origin in the latter case. Interestingly, a pronounced minimum is observed in $L(B)/L_0$ which, at sufficiently low T, occurs right at the crossover field $B^*(T)$ (cf. vertical arrows) and has the same width as the small-to-large Fermi-surface crossover (cf. horizontal bars), as inferred from Fig. 1. In an isothermal experiment that allows for the continuous registration of $L(B)/L_0$ up to B = 12 T, this minimum was clearly resolved at B ≈ 0.5 T for a temperature as high as T = 0.49 K [29]. In accordance with the initial, almost infinite slope of the $T^*(B)$ crossover line at low T (Fig. 1), the position $B^*$ of the $L(B)/L_0$ minimum at T = 0.1 K, the lowest temperature used in this study, is close to the critical field $B_N$ = 0.06 T. Here, its depth already amounts to ≈ 0.9 (see Fig. 5). Therefore, it is reasonable to assume that upon cooling, the electronic Lorenz ratio $L_{el}(B)/L_0$ will display a continuously narrowing dip of depth ≈ 0.9. At zero temperature, this will take the form of a delta function at $B^* = B_N$ [24].

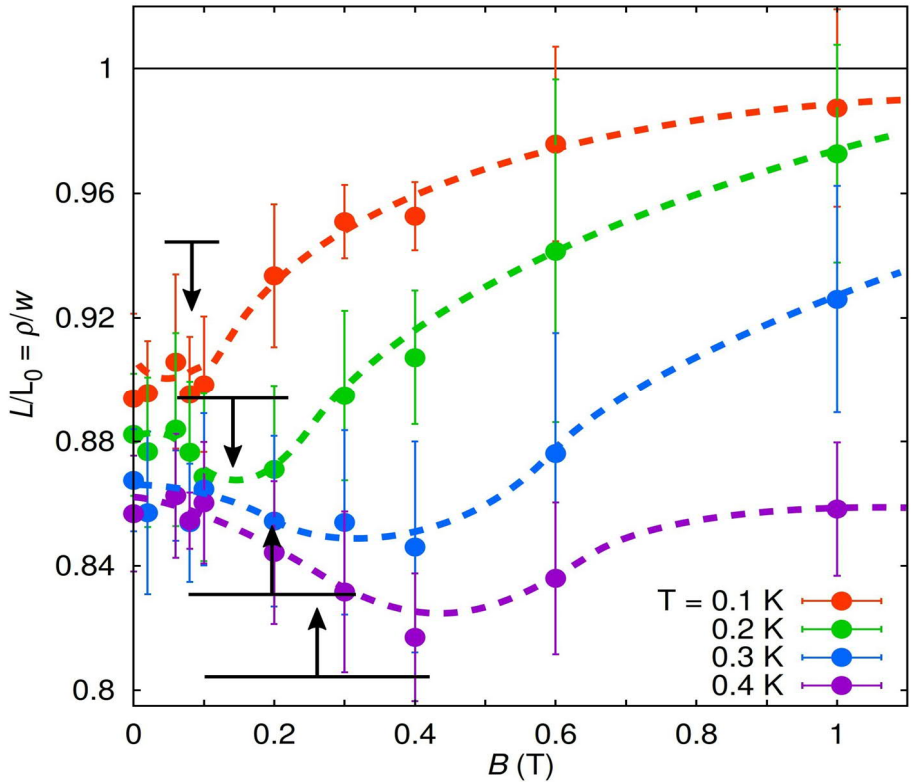

Fig. 5. Isothermal field dependences of the Lorenz ratio $L(B)/L_0$ in YbRh$_2$Si$_2$ single crystals at differing temperatures in the paramagnetic regime, 0.1 K ≤ T ≤ 0.4 K, where the thermal transport is predominantly electronic because of the evident lack of (para)magnon excitations. The data were taken from the temperature traces shown in Fig. 2. A minimum is found near the Mott-type small-to-large-Fermi surface crossover field B*(T) (vertical arrows), as displayed in Fig. 1, and it has a width similar to that of this crossover (horizontal bars). This indicates an additional inelastic scattering process attributed to fermionic quantum critical fluctuations (see text). In the zero-temperature limit, this minimum in the electronic Lorenz ratio is extrapolated to be a sharp drop exactly at the critical field $B_N$ = 0.06 T, where $L(B_N)/L_0 \approx 0.9$, indicating a violation of the WF law. Reproduced from Pfau et al. [24]

III. Quantum Critical Small-to-Large Fermi Surface Fluctuations: Prime Candidate for Driving Strange Metallicity and Violation of the WF law

How to interpret the afore-reached conclusions? First, at finite temperatures, the broadened minimum observed in the isothermal field dependence of the Lorenz ratio indicates a new inelastic scattering process intimately related to the small-to-large Fermi-surface crossover at the Kondo-breakdown QCP. Second, the small-to-large Fermi-surface fluctuations represent quantum critical fluctuations of a fermionic nature, which differ from their bosonic counterparts that involve spin fluctuations. Unlike the latter, these fermionic excitations persist at T = 0, where the small and the large Fermi surfaces coexist; for, upon cooling, their quasiparticle weights smoothly disappear at the Kondo-destroying QCP [24]. We conclude that these fermionic small-to-large Fermi-surface fluctuations serve as the primary scattering centers for the charge and heat carriers, i.e., the low-energy excitations of the two Fermi surfaces. They are, therefore, considered the prime candidate responsible for the strange metallicity and violation of the WF law in $YbRh_2Si_2$. Future theoretical work is needed to characterize the microscopic nature of this novel type of quantum critical fluctuation.

Acknowledgments

We gratefully acknowledge insightful discussions with Qimiao Si, Michael Smidman, Steffen Wirth and Huiqiu Yuan as well as stimulating conversations with Piers Coleman, Stefan Lausberg, Heike Pfau, Oliver Stockert and Ulrike Stockert.